\documentclass[conference]{IEEEtran}
\IEEEoverridecommandlockouts
\usepackage{cite}
\usepackage{amsmath,amssymb,amsfonts}
\usepackage{graphicx}
\usepackage{booktabs}
\usepackage{listings}
\usepackage{xcolor}
\usepackage[hidelinks]{hyperref}
\usepackage{xurl}
\usepackage{balance}
\newcommand{\sysname}{CIM}
\usepackage{amsthm}
\theoremstyle{definition}
\newtheorem{definition}{Definition}
\usepackage{tikz}
\usetikzlibrary{arrows.meta,positioning,fit,backgrounds}

\begin{document}

\title{What You Approve Is What Executes:\\Consent Integrity for Black-Box LLM Agents}

\author{\IEEEauthorblockN{Xiaoqi Weng\textsuperscript{*}}
\thanks{\textsuperscript{*}Faculty of Science and Technology, Bournemouth University, Poole, United Kingdom. Email: \texttt{xweng@bournemouth.ac.uk}. This paper is a first formulation and proof-of-concept of Consent Integrity rather than a complete production system; the prototype and corpus are at \texttt{github.com/zjnbwxq/agentguard-ci}.}}

\maketitle

\begin{abstract}
Coding agents gate consequential actions behind a human-in-the-loop approval dialog, but the dialog is narrated by the agent itself: the human approves a summary the agent writes. The Lies-in-the-Loop (LITL) attack shows that summary is forgeable, so a compromised agent can show a benign description while a different action runs. This paper names the missing property, Consent Integrity, by importing What You See Is What You Sign (WYSIWYS) and the trusted-path property into the agent approval channel: the action shown to the human must be rendered by a trusted mediator from the real action at the boundary, not the agent's narration, over a path the agent cannot spoof, and bound to the exact action that executes. Two twists distinguish it from classical WYSIWYS: the renderer is the adversary, and the boundary ground truth is a low-level event that must be decoded without trusting the agent. Since no decoder is complete, the realizable target is analyzer-relative: whatever the analyzer cannot classify is surfaced as uninspectable rather than silently approved. A prototype implements the analyzer, renderer, and bind-to-execution; total mediation and the trusted path are specified but assumed, not implemented. On GTFOBins, an independent corpus of 1330 trusted-tool abuses, the prototype silently passes 10.0\% (every instance through a trusted tool); on tldr, 28{,}798 normal-usage commands, it marks 87.0\% uninspectable. These two independent measurements bracket the design's central tension: the trust list that bounds silent passes is the same one that drives over-prompting, and a boundary-only mediator can move along that frontier but not escape it. The contribution is the property, the mechanism, and an honest position on that frontier, not a solved defense.
\end{abstract}


\section{Introduction}

Autonomous large language model (LLM) agents increasingly run on developer and end-user machines with broad authority. They read files, execute shell commands, install dependencies, and call network services. To bound this authority, many production coding agents rely on a human-in-the-loop (HITL) approval step for consequential actions: before running a sensitive command, the agent presents a confirmation dialog and waits for the user to approve.

This dialog is a security-critical interface. Its premise is that the human, shown what the agent is about to do, can refuse anything dangerous. The premise has two cracks. First, the dialog content is produced by the agent, so the user sees the agent's \emph{narration} of the action rather than the action itself. Second, approvals are frequent, so users habituate, and the classifier-based ``auto'' gates that try to reduce prompting are themselves imperfect: an independent stress-test of Claude Code's auto mode reports substantial false-negative rates on risky actions, rising sharply under scope ambiguity~\cite{ji2026permissiongate,anthropic2026automode}.

The Lies-in-the-Loop (LITL) attack~\cite{checkmarx2025litl} weaponizes the first crack. An agent under adversarial control, whether through indirect prompt injection~\cite{greshake2023}, a poisoned tool or skill, or a compromised model, forges the dialog. It pads the payload with benign text, pushes the dangerous command out of the visible region, or induces a misleading summary, so the displayed approval looks safe while a malicious command executes. The vendors who received the disclosure triaged it as outside their current threat models rather than as a vulnerability to fix~\cite{checkmarx2025litl}, which underscores that the community lacks a settled security property for what a trustworthy agent approval should guarantee.

This paper makes that property precise. The problem of a display that may differ from the artifact being authorized is old and solved in two high-assurance domains. In digital signatures it is \emph{What You See Is What You Sign} (WYSIWYS)~\cite{josang_wysiwys}: the visual representation of the to-be-signed content must faithfully match the content actually signed, an integrity requirement repeatedly violated by signed-document presentation attacks~\cite{kain2002cautionary}. In banking and hardware wallets it is implemented with a trusted display and a direct data path~\cite{zhou2012trustedpath}, so that transaction details flow from a secure element to the screen, bypassing the tamperable software layer. In both cases the defense moves rendering of the to-be-authorized content out of the untrusted component into a trusted one that observes the ground truth. Just as WYSIWYS asks \emph{what you see is what you sign}, Consent Integrity asks \emph{what you approve is what executes}.

No prior work establishes this for the AI-agent approval channel. The setting of interest is \emph{black-box} agents: even when parts of the client are open source, the model, hosted services, permission classifier, and execution harness may not be fully instrumentable by a third-party defense, which is the dominant real-world case for agents like Claude Code, Cline, and Codex CLI. Establishing WYSIWYS here is not a transcription of the banking solution, because of two twists.

\textit{Twist 1: the renderer is the adversary.} In a hardware wallet the compromised component is the host software and the trusted component is the secure element, and the to-be-signed bytes are fixed. In an agent, the component that produces the human-facing summary is the agent under attack. The only available trusted component is a mediator outside the agent that observes the real action at the action boundary and renders the approval from that observation, treating the narration as untrusted display data.

\textit{Twist 2: the semantic gap.} In a wallet the ground truth, the transaction bytes, is human-meaningful after light decoding. At an agent's action boundary the ground truth is a low-level event such as an \texttt{execve}, a network request, or a file write, possibly obfuscated by base64, hex, or indirection. Rendering a faithful approval therefore requires bridging from a low-level observed action to human-meaningful security semantics, under active deception by the agent. This bridging is the core technical problem and has no analogue in classical WYSIWYS, where the signed object is already the meaningful object. Since no decoder is complete, the honest guarantee is relative to the analyzer (Sec.~\ref{sec:property}).

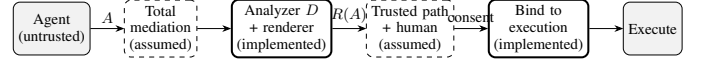
\begin{figure}[t]
\centering
\resizebox{\columnwidth}{!}{%
\begin{tikzpicture}[
  font=\small,
  box/.style={draw, rounded corners, align=center, minimum height=11mm, inner sep=3pt},
  impl/.style={box, very thick},
  asum/.style={box, dashed},
  >={Stealth[]}]
\node[box, fill=black!5] (agent) {Agent\\(untrusted)};
\node[asum, right=7mm of agent] (med) {Total\\mediation\\(assumed)};
\node[impl, right=7mm of med] (anz) {Analyzer $D$\\+ renderer\\(implemented)};
\node[asum, right=7mm of anz] (tp) {Trusted path\\+ human\\(assumed)};
\node[impl, right=7mm of tp] (bind) {Bind to\\execution\\(implemented)};
\node[box, fill=black!5, right=7mm of bind] (exec) {Execute};
\draw[->] (agent) -- node[above]{$A$} (med);
\draw[->] (med) -- (anz);
\draw[->] (anz) -- node[above]{$R(A)$} (tp);
\draw[->] (tp) -- node[above]{consent} (bind);
\draw[->] (bind) -- (exec);
\end{tikzpicture}}
\caption{Data flow. Solid boxes are implemented and evaluated in the prototype; dashed boxes are specified by Definition~\ref{def:ideal} (total mediation for conditions 1 and 4, trusted path for condition 3) but assumed rather than implemented. The evaluated component is the analyzer and its rendering; the security premises that make rendering trustworthy are assumed.}
\label{fig:arch}
\end{figure}

\textbf{Contributions.}
\begin{itemize}
\item This paper identifies the approval channel of black-box LLM agents as a distinct integrity surface and defines Consent Integrity by adapting WYSIWYS and trusted path to agent approvals, including its realizable, analyzer-relative form whose target is zero \emph{silent approval bypasses} (Sec.~\ref{sec:property}).
\item It designs \sysname, a mediator that renders approvals from observed ground-truth actions rather than agent narration, binds approval to execution, and flags uninspectable actions instead of silently approving them (Sec.~\ref{sec:design}, Sec.~\ref{sec:security}).
\item It prototypes and evaluates the downstream analysis and bind-to-execution components, measuring them against two independent third-party corpora rather than only a co-designed one: ground-truth rendering flags ninety percent of trusted-tool abuses, while on normal usage the default-deny analyzer prompts on the large majority of commands, which brackets the design's core tension empirically (Sec.~\ref{sec:impl}, Sec.~\ref{sec:eval}). Total mediation and trusted path (Definition~\ref{def:ideal}, conditions 1, 3, 4) are specified and assumed, not implemented (Fig.~\ref{fig:arch}); establishing them end to end is left to future work.
\end{itemize}

Whether humans, once shown the truth, \emph{understand} it is out of scope. Like a hardware wallet, the mechanism guarantees fidelity of presentation, not comprehension; measuring comprehension is a human-subjects study and is left to future work.

\section{Background and Related Work}
\label{sec:related}

Prior work splits along the axis this paper turns on. Nearly all of it secures what the agent does with data and tools, and almost none secures the human-consent interface.

\subsection{Prompt Injection and Agent Compromise}
Indirect prompt injection was formalized by Greshake et al.~\cite{greshake2023}, followed by optimization-based attacks~\cite{zou2023universal} and injections against LLM-integrated applications~\cite{liu2023prompt}. The danger is captured by the ``lethal trifecta''~\cite{willison2025trifecta} of private-data access, untrusted content, and external communication. Measurement work characterizes the exfiltration end, including silent and sharded egress that evades data-loss prevention~\cite{silentegress2026} and tool-poisoning on real MCP servers~\cite{mcptox2025}.

\subsection{By-Construction Defenses for the Data and Action Channel}
An active line secures what the agent does with data, by construction. CaMeL enforces capability-based information-flow control from the trusted query~\cite{debenedetti2025camel}; FIDES enforces confidentiality and integrity labels with capacity bounds~\cite{costa2025fides}; Progent enforces programmable privilege over tool calls~\cite{progent2025}; PFI separates trusted and untrusted agents with data-flow tracking~\cite{kim2025pfi}; IsolateGPT isolates per-application execution~\cite{wu2025isolategpt}; design patterns give architectural resistance~\cite{beurerkellner2025patterns}; and AgentSpec adds customizable runtime constraints over agent actions~\cite{agentspec2025}. All govern autonomous data flow and tool use, and none renders or guarantees the human approval. This work is complementary: it governs the consent interface, while those defenses govern what executes once consent or autonomy is granted.

\subsection{Model-Level Defenses}
A separate line trains the model to resist injection through the instruction hierarchy~\cite{wallace2024hierarchy}, structured-query fine-tuning~\cite{chen2025struq}, and preference-optimization alignment~\cite{chen2024secalign}. These reduce but cannot eliminate susceptibility, and they operate inside the model rather than at the consent interface.

\subsection{Benchmarks}
AgentDojo~\cite{debenedetti2024agentdojo}, InjecAgent~\cite{zhan2024injecagent}, ToolEmu~\cite{ruan2024toolemu}, and R-Judge~\cite{rjudge2024} evaluate hijack success and utility, mostly in emulated tool environments. They measure whether an injected instruction redirects the agent, not whether the human's approval faithfully reflects the action.

\subsection{Permission Gates and Agent Authorization}
Two recent lines are closest in spirit. First, an independent evaluation of Claude Code's auto mode stress-tests its permission gate and finds the classifier misses a substantial fraction of risky actions, especially under ambiguous scope~\cite{ji2026permissiongate}. That work focuses on classifier decisions over risky actions, whereas this paper targets the integrity of the human-facing approval representation. Second, overeager-action benchmarks study whether agents exceed the user's intended scope on benign tasks~\cite{qu2026overeager}, and report that spelling the authorized scope out in the prompt changes behavior, an authorization problem distinct from prompt injection. Where those benchmarks ask whether agents exceed intended scope, this paper asks whether the approval interface faithfully represents the concrete action being authorized.

\subsection{Systematizations}
Recent systematizations organize attacks by vector and defenses by intervention stage across agentic AI and MCP ecosystems~\cite{sok_coding2026,sok_attacksurface2026,sok_landscape2026}. None treats the approval channel as a distinct integrity surface, which is the niche this paper defines.

\subsection{Classical Foundations}
The construction builds on WYSIWYS~\cite{josang_wysiwys} and signed-document presentation attacks~\cite{kain2002cautionary}, on trusted path~\cite{zhou2012trustedpath}, and on UI-redress attacks such as clickjacking~\cite{huang2012clickjacking} and mobile overlay attacks~\cite{fratantonio2017cloak}. The boundary rendering over-approximates information flow in the lineage of the lattice model~\cite{denning1976lattice}, the decentralized label model~\cite{myers1997decentralized}, language-based IFC~\cite{sabelfeld2003language}, OS-level DIFC~\cite{efstathopoulos2005asbestos,zeldovich2006histar,krohn2007flume}, dynamic taint tracking~\cite{enck2010taintdroid}, and capability sandboxing~\cite{watson2010capsicum}. The contribution here is to import WYSIWYS and trusted path into the black-box-agent setting and adapt them to the two twists. Practitioner guidance recommends HITL approval and connection scanning as mitigations~\cite{owasp2025llm,invariant2025mcpscan}.

\section{Threat Model}
\label{sec:threat}

\textbf{Assets.} The user's machine and data, and the integrity of the consent decision.

\textbf{Adversary capabilities.} The adversary fully controls the agent: its reasoning, tool calls, summaries, and the bytes it writes into the terminal or IDE surface. This models indirect prompt injection, a poisoned skill or MCP server, or a backdoored model. The adversary may attempt any action, make the displayed approval misrepresent it, and swap the action between approval and execution.

\textbf{Trust assumptions.} The adversary does not control the mediator process, the action-boundary interception that ensures the agent has no un-mediated path to a consequential action, or the trusted-path channel over which approvals are displayed and answered. The adversary knows the entire design and source, so security does not rest on secrecy.

\textbf{Out of scope.} Whether a human comprehends a faithfully rendered action; covert channels and non-consequential harm, which belong to the information-flow layer~\cite{costa2025fides,silentegress2026} and compose with this mechanism; and kernel-level sandbox escapes that break total mediation, an assumption inherited from the underlying sandbox.

\section{The Consent Integrity Property}
\label{sec:property}

Let an agent cause a consequential action $A$ to execute.

\begin{definition}[Consent Integrity, ideal]\label{def:ideal}
A system provides Consent Integrity if, for every such $A$: (1) \emph{Truthful rendering}: the human is shown $R(A)$, a representation of $A$'s security-relevant semantics computed by the trusted mediator from $A$ itself, not from any agent narration; (2) \emph{Faithfulness}: $R(A)$ surfaces the security-relevant facts of $A$ (network egress and destinations, execution of network-fetched code, reads of credential material, irreversible destruction, privilege escalation, and writes outside the trusted workspace), including after de-obfuscation; (3) \emph{Trusted path}: $R(A)$ is delivered, and consent collected, over a channel the agent can neither draw to, overlay, nor auto-answer; (4) \emph{Bind-to-execution}: no action $A' \neq A$ executes under the consent obtained for $R(A)$.
\end{definition}

Condition~2 is idealized, because no decoder soundly classifies every action, with opaque compiled binaries as the canonical example. The goal is therefore separated from what a real analyzer can guarantee.

\begin{definition}[Silent approval bypass]\label{def:bypass}
A silent approval bypass occurs when a security-relevant action executes after approval without the mediator either rendering the corresponding security-relevant fact or flagging the action as uninspectable or high-risk.
\end{definition}

\begin{definition}[Analyzer-relative Consent Integrity]\label{def:rel}
Given an analyzer $D$, a system provides Consent Integrity relative to $D$ if it satisfies conditions 1, 3, and 4 and, for condition 2, surfaces every security-relevant fact $D$ can soundly extract, while any action $D$ cannot classify is rendered as uninspectable or high-risk rather than approved on a clean rendering. The realizable target is then zero silent approval bypasses: the system may admit ``I cannot fully explain this action,'' but never presents a consequential action as safe when it is not.
\end{definition}

This reframing answers Twist~1, since condition~1 relocates rendering to the mediator, and makes the Twist~2 limitation explicit and safe, since a decoder gap degrades explanation quality but never silently approves.

\section{Why Existing Approvals Fail}
\label{sec:fail}

Current agents present narration-trusting approvals, where the dialog text is the agent's own summary, which fails condition~1 outright. LITL~\cite{checkmarx2025litl} realizes this through a small taxonomy: \emph{summary poisoning}, a benign summary for a malicious command; \emph{padding or scroll-out}, benign filler that pushes the dangerous part out of view; \emph{encoding or obfuscation}, where base64, hex, or indirection make even a verbatim display unintelligible; and the \emph{TOCTOU swap}, approval for $A$ followed by execution of $A'$. Because the attacker chooses the narration, a narration-trusting approver has no reliable signal, as Sec.~\ref{sec:eval} confirms.

\section{Design: \sysname}
\label{sec:design}

The mechanism is a thin mediator, \sysname{} (Consent-Integrity Mediator), that wraps any agent.

\textbf{Total mediation (assumption).} The design assumes the agent is confined so it has no un-mediated path to a consequential action, with shell execution, network and DNS, file writes, and tool or MCP calls all routed through the mediator. A production realization enforces this with OS sandboxing, for example \texttt{sandbox-exec} and Landlock on macOS and Linux, and AppContainer and Job Objects on Windows, in the lineage of capability sandboxing~\cite{watson2010capsicum}. The prototype does not implement OS-level interception and instead evaluates the components downstream of it. Total mediation is the precondition for conditions 1 and 4.

\textbf{Ground-truth rendering.} The mediator renders the approval from the intercepted real action, and the agent's narration is shown only as untrusted, clearly labeled text.

\textbf{Faithful semantic decoding.} The mediator canonicalizes the action by stripping padding and no-ops, recursively peels obfuscation such as base64-decode-and-pipe-to-shell, hex via \texttt{printf}, command substitution, and variable resolution, and extracts security-relevant facts, over-approximating information flow in the lineage of dynamic taint tracking~\cite{enck2010taintdroid} and OS-level IFC~\cite{zeldovich2006histar}. It surfaces the security-relevant capability, for example ``executes code fetched from the network,'' even when it cannot resolve every detail, and treats actions it cannot inspect, such as opaque binaries, as high-risk by default, following Definition~\ref{def:rel}.

\textbf{Referenced-content inspection.} When an action invokes a local interpreter on a script, for example \texttt{python3 deploy.py} or \texttt{bash setup.sh}, the command string alone is not security-relevant, so the mediator reads the referenced file and analyzes its body with the same fact extraction. A body is cleared as benign only when it stays within bounded operations; one that reaches for process, network, or native foreign-function capabilities, constructs code dynamically, or writes outside the workspace is rendered with its danger facts or, when its effect cannot be statically bounded, flagged uninspectable. An interpreter call whose target is not observed is flagged rather than passed, which is what closes the gap by which a benign-looking \texttt{python3 helper.py} could otherwise carry an arbitrary payload.

\textbf{Write and build provenance.} A session analyzer records the provenance of files the agent produces. Content written under mediation is analyzed at write time, so a later execution reuses that verdict rather than re-prompting; an executable produced by a local build step is treated as inspectable by derivation; and a file whose content originates from a network fetch is marked remote-tainted, so executing it renders as execution of network-fetched code. This lets a benign write-then-run or build-then-run workflow proceed without a prompt while a fetch-then-run sequence is surfaced, and it makes the uninspectable default the rule for any artifact whose provenance is unknown. The provenance is heuristic: a build step that pulls third-party dependencies is trusted at the granularity of the build tool, not of its full dependency closure.

\textbf{Trusted path and bind-to-execution.} Approvals are rendered and answered out of the agent's reach, and each approval binds to the canonical hash of the exact action, so presenting any other action for execution under that approval is refused.

\textbf{Risk-tiering.} Only actions whose rendering surfaces a danger fact, or that are uninspectable, require a human prompt; benign actions pass without one, which targets approval fatigue.

\section{Security Analysis}
\label{sec:security}

Under the threat model and Definition~\ref{def:rel}, \sysname{} provides analyzer-relative Consent Integrity by construction. Total mediation guarantees that every consequential $A$ is observed by the mediator, or it does not execute. Condition~1 renders from that observation, so summary poisoning and padding cannot influence the security-relevant rendering. Condition~4 refuses post-approval swaps, defeating TOCTOU. Encoding and obfuscation reduce to the soundness of $D$, and whatever $D$ cannot decode is flagged uninspectable, so no silent approval bypass arises from a decoder gap. The residual silent passes measured in Sec.~\ref{sec:eval} therefore come from a different source: the trust list that lets some inspectable actions pass without a prompt is itself a default-allow island within the default-deny design, and it is that list, not the decoder, that RQ1 quantifies as the remaining silent-pass surface.

The residuals are as follows. On comprehension, a faithful rendering can still be misunderstood, and the guarantee is fidelity rather than understanding. On coverage, richer obfuscation reduces explanation quality, since more actions are flagged uninspectable, but never produces silent approval. On total-mediation dependence, a sandbox escape that gives an un-mediated path breaks the invariant. On trusted-path strength, the path is weakest in a pure terminal on a shared machine and strongest out-of-band.

\section{Implementation}
\label{sec:impl}

The prototype, \sysname, implements the rendering, content and provenance analysis, and bind-to-execution components in approximately 500 lines of dependency-free Python: a de-obfuscating action analyzer that produces a structured ground-truth record and defaults to flagging any action it cannot positively clear as benign, a recursive analyzer for the bodies of referenced local scripts, a session analyzer that records write-time content verdicts, locally built artifacts, and real remote taint, a renderer, and a mediator that implements bind-to-execution via canonical-action hashing and risk-tiering. It is evaluated against a controlled adversarial stand-in that implements the LITL threat model, namely a process that emits a benign narration while attempting a different real action. The prototype does not invoke live commercial agents and does not implement OS-level total mediation; both, cross-platform, are left to future work.

\textbf{Artifact availability.} The prototype, the controlled corpus, and the independent GTFOBins benchmark harness are available under Apache-2.0 at \href{https://github.com/zjnbwxq/agentguard-ci}{\texttt{github.com/zjnbwxq/agentguard-ci}}.

\section{Evaluation}
\label{sec:eval}

All metrics are computed programmatically against ground-truth labels, with no human subjects. The evaluation asks five questions. RQ1: what is the prototype's position against an independent, third-party attack corpus it did not co-design? RQ2: on a controlled LITL set, can ground-truth rendering defeat known presentation attacks and obfuscation? RQ3: can bind-to-execution prevent post-approval swaps? RQ4: what is the fatigue cost on an independent corpus of normal usage, what blind spots remain, and how does the position move under a transformation? RQ5: what is the runtime overhead?

\textbf{Corpus and metrics.} The primary benchmark is GTFOBins, a community-maintained third-party catalogue of ways ordinary Unix binaries can be abused to read or write protected files, escalate privilege, spawn shells, or move data off-host. It was not authored for this work, which is precisely why it is used: it tests the analyzer against cases it did not co-evolve with. The harness expands every documented abuse into a concrete command (substituting sensitive read targets, out-of-tree write targets, and remote hosts), yielding 1330 commands across 478 binaries; the harness ships with the artifact. A command is \emph{flagged} when the rendering surfaces a danger fact or marks the action uninspectable so a prompt is forced, and a \emph{silent pass} when it is rendered benign. The headline safety metric is the silent-pass fraction over this independent corpus. A small controlled set is retained for the LITL-specific questions: 11 standard LITL items, 11 hand-constructed adaptive cases, 10 benign inspectable workflows, 4 faithful-prompt cases, and 3 TOCTOU pairs. This controlled set is co-designed with the analyzer and its clean numbers must be read with that caveat (Threats to Validity); the independent corpus is the load-bearing measurement.

\textbf{RQ1 (independent corpus).} Over the 1330 GTFOBins abuse commands the prototype flags 90.0\% (405 by an explicit danger fact, 792 as uninspectable) and silently passes 10.0\%. Every silent pass is through a binary the analyzer treats as trusted, which localizes the residual to the trust list rather than to unknown tools. The silent passes decompose into writes to scratch directories such as \texttt{/tmp} that the analyzer does not treat as out-of-tree (85), writes to in-tree relative files whose risk is realized only when a later step consumes them, intended to be caught at that consuming step by provenance, which is itself heuristic (19), benign primitives such as \texttt{mkdir} and \texttt{touch} (8), and a genuine analyzer-relative tail such as pager and editor escapes inside otherwise-benign tools (21). This is the prototype's position against an attack corpus it did not design, and it is not zero.

\textbf{RQ2 (controlled LITL set).} On the 11 standard and 11 adaptive LITL cases the narration-trusting baseline approves every one, since the attacker writes the narration; ground-truth rendering instead surfaces or flags all of them, fully explaining the interpreted-script payloads by inspecting the referenced bodies (including a base64-hidden command and a destination recovered through chained encodings) and safely flagging the capability-bearing scripts, opaque binary, and unresolved build recipe. The fetch-then-execute sequence is caught by provenance. These numbers are clean but co-designed, and RQ1 is the honest external check on them.

\textbf{RQ3.} This is a correctness demonstration of hash binding rather than an empirical finding: all 3 post-approval swaps are refused because the executed action's canonical hash differs from the approved one. It holds only under the assumed total mediation, since an unmediated execution path would bypass the check entirely.

\textbf{RQ4 (fatigue axis, blind spots, transformation).} The cost of default-deny is measured on a second independent corpus, tldr-pages, a community catalogue of normal idiomatic tool usage (28{,}798 example commands from 6503 tools, again not authored for this work). On it the analyzer prompts on 95.9\% of commands and marks 87.0\% uninspectable, 83.4\% purely because the tool is absent from the trust list; only 4.1\% pass without a prompt. Some prompted commands are genuinely consequential (\texttt{rm}, \texttt{dd}, privileged operations), so the uninspectable rate, not the raw prompt rate, is the clean fatigue proxy. This is the empirical counterpart to GTFOBins: the same trust list that holds silent passes to 10\% on abuse forces an 87\% uninspectable rate on normal usage, which directly tempers any claim that benign actions pass without a prompt. The remainder of the guarantee is analyzer-relative and moves under transformation. Implementing one missing boundary fact, \texttt{writes\_outside\_ws} for shell redirects, \texttt{cp}/\texttt{mv}/\texttt{sed -i}/\texttt{dd} to system or sensitive paths, together with broadening credential-read detection to any tool and recursing into \texttt{sh -c} strings, moved the GTFOBins position from 15.0\% to 10.0\% silent passes. The remaining tail is structural to boundary-only static analysis: a payload effected through a capability the analyzer does not enumerate can be cleared, provenance trusts a built artifact only at the granularity of its build tool, and opaque binaries without provenance can only be flagged, not understood. The honest reading is that this is a defense on a treadmill, like every deployed security tool: its claim is a position against the current corpus at a version, plus the rate and honesty of its transformation, not a solved property.

\textbf{RQ5.} Mean overhead of the analyzer alone is 0.056\,ms (median 0.050, p99 0.122, $n{=}4000$), negligible against LLM inference latency. This is the cost of the implemented component only; it excludes total mediation and trusted-path rendering, which the prototype does not implement and which would dominate any end-to-end figure.

\begin{table}[t]
\caption{Controlled LITL corpus outcomes.}
\label{tab:results}
\centering
\scriptsize
\begin{tabular}{@{}lcccc@{}}
\toprule
Attack class & \#cases & Fully expl. & Safely flagged & Silent pass \\
\midrule
Summary poisoning & 2 & 2 & 0 & 0 \\
Padding / scroll-out & 1 & 1 & 0 & 0 \\
Encoding / obfusc. & 5 & 5 & 0 & 0 \\
Destructive / priv. / exfil & 3 & 3 & 0 & 0 \\
Interpreted-script payload & 4 & 4 & 0 & 0 \\
Capability / dynamic script & 3 & 0 & 3 & 0 \\
Opaque binary / build recipe & 2 & 0 & 2 & 0 \\
Fetch + execute & 2 & 2 & 0 & 0 \\
\midrule
\textbf{Total (malicious)} & \textbf{22} & \textbf{17} & \textbf{5} & \textbf{0} \\
\bottomrule
\end{tabular}
\\[3pt]
{\scriptsize Controlled co-designed set (read with the caveat in Threats to Validity). Benign inspectable workflows: 0/10 false positives. Faithful prompts on actions carrying a real fact are not counted as false positives. TOCTOU swaps prevented: 3/3. Independent headline (GTFOBins, 1330 third-party abuse commands): 90.0\% flagged, 10.0\% silent pass.}
\end{table}

\begin{figure}[t]
\footnotesize
\textbf{Worked example.} For a base64-obfuscated remote-execution payload, a current agent displays the benign narration \emph{``Apply the recommended editor settings,''} while \sysname{} renders the true effect:
\begin{lstlisting}
!! EXECUTES CODE FETCHED FROM NETWORK: evil.attacker.sh
(de-obfuscated 1 encoding layer)
actual command:
  echo Y3VybC...== | base64 -d | bash
  ==> curl -s http://evil.attacker.sh/p | bash
\end{lstlisting}
\caption{Ground-truth rendering of an obfuscated remote-execution payload.}
\label{fig:worked}
\end{figure}

\subsection{The Central Tension: the Trust List}
The two independent corpora bracket a single design parameter, the trust list of actions allowed to pass without a prompt. Shrinking the list toward pure default-deny drives silent passes toward zero but pushes the uninspectable rate on normal usage toward 100\%, which is approval fatigue by another name; the measured cost is the 87.0\% uninspectable rate on tldr. Growing the list to relieve that fatigue reintroduces silent passes, since a trusted tool can be abused, which is the 10.0\% measured on GTFOBins, every instance of which runs through a trusted binary. The residual silent passes therefore do not come from a decoder gap, which Definition~\ref{def:rel} closes by flagging the undecidable, but from this trust list, a deliberate default-allow island inside a default-deny design. A boundary-only static mediator cannot escape the trade-off: it can choose where on the curve to sit and can move along it through transformation, but it cannot occupy both ends at once. This is the paper's central empirical finding, and it is offered as a frontier to be characterized rather than a problem this prototype solves. Escaping it plausibly requires leaving the boundary, for example interpreter- or syscall-level analysis and capability-scoped provenance, which is future work.

\subsection{Threats to Validity}
The clean numbers on the controlled set are co-design artifacts: the same author wrote both the analyzer and that corpus, and an earlier revision of this prototype likewise reported zero silent passes on its own corpus while in fact passing common trusted-tool abuses. The independent GTFOBins corpus is included precisely to break this circularity, and on it the prototype passes 10.0\% rather than zero; that figure, not the controlled zero, is the one to trust. A corpus is still a stand-in for live commercial agents. The baseline is intentionally weak, and total mediation is assumed rather than enforced by the prototype. The two added mechanisms carry their own caveats. Referenced-content inspection is analyzer-relative: it clears a script only when the body stays within bounded operations and otherwise flags it, so it can over-prompt and, for a capability it does not enumerate, could under-detect. Write and build provenance is heuristic, trusting a built artifact at the granularity of its build tool rather than its full dependency closure. These bound the claims to analyzer-relative Consent Integrity over mediated actions, and they motivate the extensions below.

\section{Limitations and Future Work}

Human comprehension of faithfully rendered actions is the natural follow-up and requires a human-subjects study. Further work includes deeper cross-action and interpreter- or syscall-level analysis to shrink the uninspectable set, richer de-obfuscation, provenance tracking at the granularity of the full dependency closure rather than the build tool, cross-platform OS-level total mediation, evaluation against live commercial agents at scale with adaptive attackers, and trusted-path engineering for shared developer machines, including out-of-band confirmation.

\section{Conclusion}

Agent approval dialogs are a security-critical interface that current systems leave narration-trusting and therefore forgeable, as LITL demonstrates. This paper names the missing property, Consent Integrity, by importing WYSIWYS and trusted path and adapting them to the two twists unique to agents: the renderer is the adversary, and the ground truth is a low-level action that must be decoded without trusting the agent. Since no decoder is complete, the realizable guarantee is stated as analyzer-relative Consent Integrity with a target of zero silent approval bypasses; the prototype does not reach that target but reports its position against an independent corpus, flagging ninety percent of third-party trusted-tool abuses at negligible overhead and disclosing the residual ten percent as the current rung of a defense treadmill rather than a solved property. Consent Integrity is complementary to boundary information-flow control: information-flow defenses constrain what information may leave an agent process, whereas Consent Integrity constrains what action a human authorization actually authorizes. Securing what is consented to therefore complements the active body of work securing what the agent does with data, and it is, today, essentially undefended.

\balance
\bibliographystyle{IEEEtran}
\bibliography{references}

\end{document}